\documentclass[prl,aps,twocolumn, superscriptaddress, floatfix, nofootinbib,preprintnumbers,10pt]{revtex4-1}
\usepackage{graphicx}
\usepackage{latexsym}

\usepackage{ulem}
\usepackage[usenames]{color}
\definecolor{DarkGreen}{rgb}{0.0,0.5,0.0}

\begin{document}

\preprint{IPMU-10-0199}
\title{Warm DBI Inflation}

\author{Yi-Fu Cai}
\affiliation{Department of Physics \& School of Earth and Space Exploration \& Beyond Center, Arizona State University, Tempe, Arizona 85287-1504, USA}
\author{James B. Dent}
\affiliation{Department of Physics \& School of Earth and Space Exploration \& Beyond Center, Arizona State University, Tempe, Arizona 85287-1504, USA}
\author{Damien A. Easson}
\affiliation{Department of Physics \& School of Earth and Space Exploration \& Beyond Center, Arizona State University, Tempe, Arizona 85287-1504, USA}
\affiliation{Institute for the Physics and Mathematics of the Universe, University of Tokyo, 5-1-5 Kashiwanoha, Chiba 277-8568, Japan}



\begin{abstract}
We propose a warm inflationary model in the context of relativistic D-brane inflation in a warped throat, which has Dirac-Born-Infeld (DBI) kinetic term and is coupled to radiation through a dissipation term. The perturbation freezes at the sound horizon and the power spectrum is determined by a combination of the dissipative parameter and the sound speed parameter. The thermal dissipation ameliorates the {\it eta} problem and softens theoretical constraints from the extra-dimensional volume and from observational bounds on the tensor-to-scalar ratio. The warm DBI model can lead to appreciable non-Gaussianity of the equilateral type. As a phenomenological model, ignoring compactification constraints, we show that large-field warm inflation models do not necessarily yield a large tensor-to-scalar ratio.
\end{abstract}

\maketitle





Inflationary cosmology has become the prevalent paradigm to describe the physics of the early universe. Inflation solves the flatness problem, the homogeneity and unwanted relics problems and predicts a nearly scale-invariant primordial power spectrum consistent with current cosmological observations \cite{Guth:1980zm}. Among the most famous obstacles encountered in inflationary model building is the so-called {\it eta} problem \cite{Copeland:1994vg}. If the inflaton field has a Dirac-Born-Infeld (DBI) action \cite{Silverstein:2003hf} it is possible to circumvent the {\it eta} problem \cite{Easson:2009kk}. The special form of the DBI kinetic term introduces an effective speed limit on the inflaton field which can keep the inflaton near the top of a potential, even if the potential is steep. In string theory models of D-brane inflation, the inflaton field can be interpreted as a moduli parameter of a D-brane  \cite{Myers:1999ps} moving in a warped throat region of an approximate Calabi-Yau, flux compactification. The propagation of field fluctuations is characterized by the sound speed parameter, $c_s$, which is characteristically smaller than unity in DBI models.  Because the sound speed is slower than the speed of light, cosmological perturbations are frozen at the sound horizon where $k = aH/c_s$, rather than the Hubble radius where $k = aH$.

It was soon recognized that ``ultraviolet" (UV) DBI brane inflation is strongly constrained by the combination of backreaction effects from a relativistic brane moving in a warped throat \cite{Silverstein:2003hf, Easson:2007dh}, and from the inflationary background \cite{Chen:2008hz}. It is difficult for models to provide enough e-foldings to be compatible with cosmological data \cite{Bean:2007hc}. Moreover, the simplest models suffer from a severe inconsistency problem due to an upper bound on the tensor-to-scalar ratio arising from constraints on the throat volume \cite{Baumann:2006cd} and bulk volume \cite{Chen:2006hs}, and a lower bound from observations \cite{Lidsey:2007gq}. Successful reheating also poses a significant challenge. Typically, the reheating process involves brane antibrane annihilation and leads to the production of highly constrained, long-lived, Kaluza-Klein (KK) modes \cite{Barnaby:2004gg}.

The standard picture of inflation involves a thermodynamically supercooled phase of the early Universe due to the exponential expansion. Standard model particles are produced via the traditional reheating mechanism after inflation ends. An alternative proposal is {\it warm inflation} \cite{Berera:1995wh}, during which the vacuum energy of the inflaton is released into relativistic particles through a dissipation term during inflation, thus eliminating the necessity of a reheating process \cite{Fang:1980wi}. A distinguishing feature of this mechanism is that the {\it eta} problem is alleviated by a large dissipation parameter, since the radiation generated by the time dependence of the inflaton acts to slow the inflaton field \cite{Berera:2008ar}. Inflation ends once the initially subdominant radiation energy density approaches that of the vacuum $\rho_r \simeq \rho_{\rm v}$.

In this paper, we propose a model of warm DBI inflation inspired by the description of a D-brane coupled to radiation through a dissipation term. A slow-roll realization of the warm inflation naturally emerges in the nonrelativistic limit. We use the reduced Planck mass, $m_p^2=1/8\pi{G}=1$, and metric signature $(-,+,+,+)$.

D-brane inflation can be effectively described by the DBI action,
\begin{eqnarray}
 {\cal L}_{DBI} = f^{-1} \bigg[1-\sqrt{1-2fX}\bigg] - V~,
\end{eqnarray}
where we define $X\equiv-\frac{1}{2}g^{\mu\nu}\nabla_{\mu}\phi\nabla_{\nu}\phi$, $\phi$ is the inflaton field with potential $V$, and $f$ is the warp factor of an AdS-like throat. In the traditional (cold) DBI inflationary picture, the above action is coupled to the Einstein-Hilbert action. The process of the probe brane falling into the warped throat--from the top of the throat towards the tip--corresponds to a UV model. The inflaton equation of motion is
\begin{eqnarray}
 \frac{d}{dt}\bigg(\frac{\dot\phi}{c_s}\bigg) +3H\frac{\dot\phi}{c_s} -\frac{c_s}{a^2}\vec{\nabla}^2\phi +\frac{f'}{f^2}(1-c_s)+V' =0,
\end{eqnarray}
where the dot denotes the derivative with respect to cosmic time $t$, and the prime denotes partial differentiation with respect to $\phi$. In the above, we have introduced the sound speed parameter $c_s={(1-f\dot\phi^2)^{1/2}}$, whose variation during inflation can be characterized by the dimensionless parameter
\begin{eqnarray}
 s=-2m_p^2\frac{c_s'}{c_s}\frac{H'}{H}~,
\end{eqnarray}
which is a newly defined slow-roll parameter.  DBI models are also characterized by the familiar slow-roll expressions,
\begin{eqnarray}
 \epsilon=2c_sm_p^2\bigg(\frac{H'}{H}\bigg)^2~,~~
 \eta=2c_sm_p^2\frac{H''}{H}~.
\end{eqnarray}
The above parameters are expected to be much less than unity during cold DBI inflation in order to give sufficient expansion and to account for the observed primordial power spectrum. Under the slow-roll assumption, the homogeneous part of the background equation of motion simplifies to $3H\dot\phi+c_sV'\simeq0$.

We extend the traditional model by introducing a dissipation term $\Gamma$. This term can be obtained from finite temperature field theory \cite{Gleiser:1993ea}, in a model of intermediate particle decay \cite{Berera:2002sp}, or in a cosmological system of a thermal brane \cite{Dvali:1999tq}. Assuming the interaction between the inflaton and radiation is indirect~\footnote{For example, to embed warm inflation into a simple supersymmetric model, one could introduce a superpotential containing the interaction $W = g_1 \Phi Y^2 + g_2 Y Z^2$, where the inflaton $\phi$ refers to the bosonic component of the superfield $\Phi$. In this case, one obtains a light superfield $Z$ and a heavy superfield $Y$ due to the coupling to $\phi$. Thus one can integrate out $Y$ and obtain an effective dissipative term for $\phi$ coupling to radiation in terms of $Z$. This kind of interaction terms can be obtained in supersymmetric models inspired by string theory \cite{BasteroGil:2009gh}.}, the dominant background equation becomes
 $(3H+\Gamma)\dot\phi+c_sV' \simeq 0~,$
after taking into account the radiation damping effects on the evolution of the inflaton field. The continuity equation for the radiation is derived to be,
 $\dot\rho_r+4H\rho_r = \Gamma\frac{\dot\phi^2}{c_s}~.$
Under the slow-roll approximation, the background equations reduce to
\begin{eqnarray}\label{sol}
 \dot\phi\simeq-\frac{c_sV'}{3H(1+\bar{\Gamma})}~,~~
 \rho_r\simeq\frac{3\bar{\Gamma}\dot\phi^2}{4c_s}~,~~
 V\simeq3m_p^2H^2~,
\end{eqnarray}
where we have introduced the dimensionless dissipative parameter, $\bar{\Gamma} \equiv \Gamma/3H$. Hence, the energy density of radiation can not be fully diluted during inflation.

Note that the consistency of the above analysis relies on behavior of a new slow-roll parameter
\begin{eqnarray}
 \sigma=2c_sm_p^2\frac{\Gamma'}{\Gamma}\frac{H'}{H}~,
\end{eqnarray}
which quantifies the variation of the dissipation term during inflation. The model of warm DBI inflation involves four slow-roll parameters $\epsilon$, $\eta$, $s$ and $\sigma$, but the fine-tuning problem is significantly diminished with respect to
the cold inflationary model, since we only require the slow-roll parameters to be smaller than $1+\bar{\Gamma}$ opposed to unity. We note that, in general, these parameters can be quite large in a strongly dissipative system. In this warm brane inflation picture, inflation will end when $\epsilon=1+\bar{\Gamma}$ and the
radiation energy density surpasses that of the probe brane potential vacuum energy.


Now we develop the theory of cosmological perturbations in the warm DBI inflation model. Since we are working in the context of a cosmological system, the metric perturbation will be included as well as field and thermal fluctuations. Similar to the model of multibrane inflation \cite{Cai:2009hw}, the propagation of each fluctuation component has its own sound speed. For convenience, we choose the spatially flat gauge with vanishing spatial curvature; accordingly, the perturbation is given by $\zeta = H \frac{\delta\rho}{\dot\rho}$.


During inflation the energy density of radiation is subdominant, and its thermal fluctuation only contributes to entropy perturbations. Consequently, we will disregard it and focus on the field fluctuation. By making use of the background solution found in Eq.(\ref{sol}), we derive the relation between the power spectrum of the perturbation and that of the field fluctuation as,
 $P_{\zeta}\simeq\frac{(1+\bar{\Gamma})^2}{2m_p^2c_s\epsilon}P_{\delta\phi}~.$
The evolution of a field interacting with radiation satisfies a Langevin equation after introducing a stochastic thermal noise $\theta$. Applying the background solution, we find the inflaton fluctuation satisfies a generalized Langevin equation which, in Fourier space, takes the form
\begin{eqnarray}\label{Langevin}
 \ddot{u}_k+(3H+\Gamma)\dot{u}_k+\frac{c_s^2k^2}{a^2}u_k \simeq
 \theta_k~.
\end{eqnarray}
If the temperature of the universe is sufficiently high \cite{Gleiser:1993ea}, the thermal noise is assumed to be highly Markovian:
 $a^3 \langle \theta_{k_1}(t_1)\theta_{-k_2}(t_2) \rangle 
 = 2\Gamma T(2\pi)^3\delta(\vec{k}_1 - \vec{k}_2)\delta(t_1-t_2)~.$
Note that the sound speed parameter in Eq.(\ref{Langevin}) can be absorbed into the gradient term by redefinition, and consequently one can solve the perturbation equation through the Green function method developed in \cite{Hall:2003zp} which yields,
 $P_{\delta\phi}
  \simeq\frac{\sqrt{3\pi}}{2} \left(1+\bar \Gamma \right)^{\frac{1}{2}}HT~.$
Hence, the sound speed parameter affects the freezing moment of the field fluctuation but does not change the fluctuation amplitude. This behavior mimics the standard DBI inflation scenario.

In warm inflation the effect of thermal noise, which is the source of density perturbations, decreases until the fluctuation amplitude freezes in, which typically happens before cosmological horizon crossing (for the scenario where the dissipation term is temperature dependent, the amplitude does not remain constant at this freeze-out point, but instead approaches a growing mode \cite{Graham:2009bf}). The curvature fluctuation can be found on superhorizon scales (super sound horizon scales for the DBI case) and then matched to the fluctuations which froze out at the smaller thermal noise scale.  Based on the above analysis and making use of $P_{\delta\phi}$, one obtains an approximately analytic solution in the strongly dissipative limit $\bar \Gamma\gg1$, for the primordial perturbation at the time of sound horizon exit, i.e.~when $k=aH/c_s$,
\begin{eqnarray}
 P_{\zeta}\simeq\frac{\sqrt{3\pi} \,\bar \Gamma^{\frac{5}{2}}\bar T H^2}{4m_p^2c_s\epsilon}~,
\end{eqnarray}
where we have defined the dimensionless temperature-to-Hubble ratio, $\bar T \equiv
T/H$. We see that the primordial perturbation is amplified by both a large dissipation parameter $\bar \Gamma$ and a small sound speed $c_s$.
The spectral index is given by
\begin{eqnarray}
 n_s-1 \equiv \frac{d\ln{P_\zeta}}{d\ln{k}}
 \simeq \frac{1}{\bar \Gamma}
 \bigg(-\frac{3}{4}\epsilon+\frac{3}{2}\eta-\frac{9}{4}\sigma-\frac{7}{4}s\bigg).
  \label{specin}
\end{eqnarray}
The above results are in agreement with those obtained in \cite{Hall:2003zp} when $c_s=1$.


Furthermore, the amplitude of the primordial tensor perturbation is given by $P_T = \frac{2H^2}{\pi^2m_p^2}$, evaluated at horizon crossing $k=aH$. The tensor-to-scalar ratio is~\footnote{Note, depending on the sound speed, the difference in horizon crossing times of the scalar and tensor components may be significant \cite{Agarwal:2008ah}; however, for the sake of the following approximation argument we will ignore this possibility. Certainly, in the slow-roll limit the difference is negligible.}
\begin{eqnarray}
 r\equiv\frac{P_T}{P_\zeta}
 \simeq\frac{8c_s\epsilon}{30.3 \bar\Gamma^{5/2} \bar T}~,
\end{eqnarray}
which implies a modified Lyth bound: the relation between the field range $\Delta\phi$ and the number of e-foldings $\Delta {\cal N}$,
\begin{eqnarray}
 \Delta\phi = \frac{\dot\phi}{H} \Delta{\cal N}
 \simeq 2.75 r^{\frac{1}{2}} \bar\Gamma^{\frac{1}{4}} \bar T^{\frac{1}{2}} m_p \Delta{\cal N}~.
\end{eqnarray}
One evaluates $\Delta\phi \gtrsim 11 r^{\frac{1}{2}} \bar\Gamma^{\frac{1}{4}} \bar T^{\frac{1}{2}} m_p$ on observable scales with $|\Delta{\cal N}|\simeq4$ (corresponding to $1<l<100$). For sufficiently large dissipation, it is possible to have a large variation of the inflaton field even if the tensor-to-scalar ratio is unobservable. This interesting characteristic behavior of warm inflation is present
even in the non-DBI limit.

Within the context of stringy compactifications, the variation of the inflaton is bounded by the volumes of the warped throat and the compact bulk \cite{Baumann:2006cd}. As a naive estimate, we require $\Delta\phi^6<{\pi}f^{-1}m_p^2/{\rm Vol}_5$, where for a typical compactification, ${\rm Vol}_5\sim \mathcal{O}(\pi^3)$. Consequently, this yields an upper bound on $r$ as a combination of $P_{\zeta}$, $\bar \Gamma$, $\bar T$, and the number of observable e-foldings $\Delta{\cal N}$:
\begin{eqnarray}
 r < \frac{8P_{\zeta}}{\sqrt{3}\pi^{\frac{5}{2}}{\rm Vol}_5 \, \bar\Gamma \, \bar T^2\Delta{\cal N}^6}~.
\end{eqnarray}


Observable levels of non-Gaussianity, such as may be produced by the warm DBI mechanism, can help to break degeneracies between inflationary models (see, e.g.~\cite{Easson:2010zy}). To calculate the non-Gaussianity produced in the warm DBI model, it is necessary to move beyond the leading order. There are two mechanisms to generate non-Gaussian fluctuations in the model under consideration. One is the non-Gaussianity produced by nonlinear thermal noise, as discussed in \cite{Moss:2007cv}; the other is nonlinear interactions of the perturbative Lagrangian. In warm inflation, non-Gaussianity in the squeezed triangle limit is expected to be negligible.

The Lagrangian expanded to cubic order is:
 ${\cal L}_3 \supseteq \frac{a^3}{2c_s^3\dot\phi} \left(\frac{1}{c_s^2}  -  1 \right)
 \left[\delta\dot\phi^3-\frac{c_s^2}{a^2}\delta\dot\phi\nabla_i\delta\phi^2 \right]~.$
For sufficiently small $c_s$, the contribution of the cubic term in the Lagrangian dominates over that of thermal noise, and we obtain the equilateral type nonlinearity parameter $f_{NL}^{equil} \simeq \frac{1}{3}(\frac{1}{c_s^2}-1)$ as in the usual DBI inflation scenario \cite{Silverstein:2003hf}. Substituting this result into the expression for the spectral index, one may obtain a lower bound on $r$ in a UV DBI model \cite{Lidsey:2007gq}\footnote{The stringent lower bound on $r$ can be relaxed in other DBI setups without dissipation, for example, multiple-field models\cite{Huston:2008ku}.}. In our case, this bound appears as a function of $n_s$, $\sigma$, $f_{NL}$, $\bar \Gamma$, and $\bar T$,
\begin{eqnarray}
 r>\frac{8(1-n_s-\frac{\sigma}{2\bar \Gamma})}{75\bar \Gamma^{\frac{3}{2}}\bar T \sqrt{3f_{NL}}}~.
\end{eqnarray}

\begin{figure}[htbp]
\includegraphics[scale=0.7]{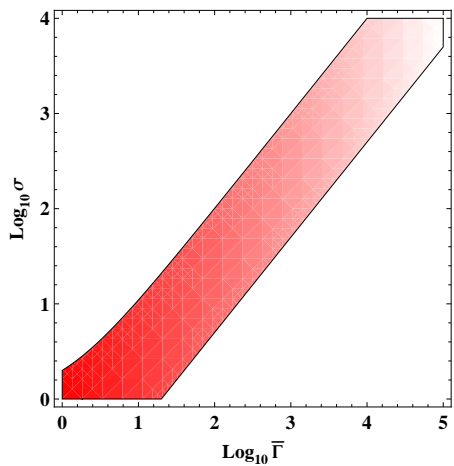}
\caption{Observational and theoretical constraints on $\bar \Gamma$ and $\sigma$.
The viable parameter space is within the red region. } \label{fig:con}
\end{figure}

By combining the upper and lower bounds we derive a lower bound on the dissipative parameter:
\begin{eqnarray}
 \frac{\bar \Gamma^{\frac{1}{2}}}{\bar T}
 >\frac{\pi^2{\rm Vol}_5 \, \Delta{\cal N}^6}{25\sqrt{3f_{NL}}\, P_{\zeta}} \left(1-n_s-\frac{\sigma}{2\bar \Gamma} \right)~.
\end{eqnarray}
Provided this inequality is satisfied, there is no inconsistency between the upper and lower bounds on $r$. This constraint is satisfied when $\sigma>2 \bar\Gamma(1-n_s)$. The latest observation of the cosmic microwave background radiation (CMB) (\it WMAP7\rm+BAO+$H_0$), yield $n_s=0.968\pm0.012$  (68\% CL) at the pivot scale $k=0.002{\rm Mpc}^{-1}$ \cite{Komatsu:2010fb}, and thus the permitted regime for $\bar\Gamma$ and slow-roll parameter $\sigma$ is presented in Fig. \ref{fig:con}. There is a color grading towards lighter tones as $\bar\Gamma$ approaches infinity, which corresponds to a fine-tuning problem possibly existing in our model.

Note that Ref. \cite{Hall:2007qw} found the $\Gamma$ parameter in slow-roll version was tightly constrained by observations. Following Section 4.4 of \cite{Hall:2007qw}, we use $\rho_r=\pi^2g_*T^4/30$ where $g_*$ is the effective partible number for radiation, and consider that the tilt of spectral index is dominated by the slow-roll parameter $\epsilon$. Making use of these approximate relations, the power spectrum of curvature perturbation can be simplified as
\begin{eqnarray}
 P_{\zeta} \simeq \frac{1}{c_s} \bigg(\frac{15}{32g_*}\bigg)^{\frac{1}{4}} \bigg(\frac{\Gamma}{m_p}\bigg)^{\frac{3}{2}} \bigg[\frac{1}{8(1-n_s)}\bigg]^{\frac{3}{4}}~,
\end{eqnarray}
which can be viewed as a generalization of Eq. (46) of Ref. \cite{Hall:2007qw}. As a consequence, we obtain a generalized constraint relation via the DBI model as
\begin{eqnarray}
 \Gamma \simeq 2^{\frac{7}{3}} 15^{-\frac{1}{6}} {g_*}^{\frac{1}{6}} (c_sP_{\zeta})^{\frac{2}{3}} (1-n_s)^{\frac{1}{2}} m_p,
\end{eqnarray}
and find it is slightly relaxed by a small $c_s$. However, a stringent constraint still exists since $c_s$ is bounded by $f_{NL}$. For typical WMAP7 data, $\Gamma \sim 10^{-7}m_p$ when $g_*\sim 100$ and $c_s\sim0.1$. Whether this constraint could bring an overproduction of gravitino or not depends on details of model construction in particle physics\cite{Sanchez:2010vj}.

The above conclusion is obtained in an extremal limit where thermal non-Gaussianity is negligible. However, away from this limit, when the non-Gaussianity may be sizeable, one finds that the sign of the estimator is usually negative which implies that the thermal fluctuation might be anticorrelated at next to leading order \cite{Moss:2007cv}. Summing these two contributions may decrease the value of $f_{NL}$ and thereby increase the viable parameter space of the model. In addition, the above constraints on $n_s$, do not allow for the possibility of running of the spectral index.  Allowing for such running significantly extends the range of the viable parameter space \cite{Cai:upcoming}.

Typically in the UV DBI model, generation of observably large non-Gaussianity is accompanied by an observably unacceptable blue spectrum \cite{Peiris:2007gz}. However, when a dissipation term is included the spectral index depends on the additional slow-roll parameter, $\sigma$ (Eq.(\ref{specin})), which, if sufficiently large, might induce a red spectral tilt. In order to determine the nature of this effect on the spectral index, one must compute all slow-roll parameters in a specific model \cite{Cai:upcoming}.

Warm inflation is capable of avoiding certain problems encountered in traditional cold inflation models (for a review see \cite{Berera:2008jn}). As we have discussed, introducing the DBI mechanism into the warm inflationary paradigm circumvents the {\it eta} problem, and opens the window for a concrete embedding of the scenario into a string theory model of D-brane inflation. The DBI form for the inflaton kinetic term offers a tantalizing opportunity to make contact with observational physics through the production of potentially measurable non-Gaussian features in the CMB. By providing a smooth transition from the inflationary phase to a radiation-dominated phase, the warm brane inflation scenario avoids the difficulties associated with reheating the post-inflationary universe via brane antibrane annihilation. As a side note we showed, for the first time, that large-field warm inflation models do not necessarily produce a large tensor-to-scalar ratio. For these reasons, we consider warm DBI inflation to be an attractive candidate to potentially enlarge the arena for inflationary model building in string theory. A rigorous construction along these lines remains a challenge for future research \cite{Cai:upcoming}.

{\sl{Acknowledgments.}} It is a pleasure to thank A. Berera, R. Flauger, T. Kobayashi, S. Mukohyama and B. Powell for useful discussions. The work of Y.F.C. and J.B.D. is supported in part by the Arizona State University Cosmology Initiative. The work of D.A.E. is supported in part by the World Premier International Research Center Initiative (WPI Initiative), MEXT, Japan and by a Grant-in-Aid for Scientific Research (21740167) from the Japan Society for Promotion of Science (JSPS), and by the Arizona State University Cosmology Initiative.



\end{document}